\documentclass[12pt]{iopart}
\usepackage{psfig}
\begin{document}

\title[Production of $a_0$-mesons in $pp$ and $pn$ reactions]
{Production of $a_0$-mesons in $pp$ and $pn$ reactions
\footnote{Supported by Forschungszentrum J\"ulich, DFG and RFFI} }

\author{E.L. Bratkovskaya \dag, V. Yu. Grishina $\|$\S,
L.A. Kondratyuk $\sharp ^+$, M. B\"uscher $^+$ and W.~Cassing \dag }

\address{\dag Institut f\"ur Theoretische Physik, Universit\"at Giessen,
     D-35392 Giessen, Germany}

\address{$\|$ Institute for Nuclear Research,
     60th October Anniversary Prospect 7A,  117312 Moscow, Russia }

\address{\S Institute of Theoretical and Experimental Physics,
     B.Cheremushkinskaya 25, 117259 Moscow, Russia }

\address{$^+$ Institut f\"ur Kernphysik, Forschungszentrum J\"ulich,
     D-52425 J\"ulich, Germany}


\begin{abstract}
We investigate the cross section for the  reaction $NN \to
NNa_0$ near threshold and at medium energies.  An effective Lagrangian
approach with one-pion exchange is applied to analyze different
contributions to the cross section for different isospin channels. The
Reggeon exchange mechanism is also considered.  The results are used to
calculate the contribution of the $a_0$ meson to the cross sections and
invariant $K \bar K$ mass distributions of the reactions  $pp\to pn
K^+\bar K^0$ and $pp\to pp K^+K^-$. It is found that the experimental
observation of $a_0^+$ mesons in the reaction $pp\to pn K^+\bar K^0$ is
much more promising than the observation of $a_0^0$ mesons in the
reaction $pp\to pp K^+K^-$.
\end{abstract}

\vspace{0.5cm}\noindent
PACS: {25.10.+s, Meson production and
13.75.-n, Proton induced reactions

\submitto{\JPG}

\maketitle

\section{Introduction}

The excitations of the QCD vacuum with different quantum numbers as well
as their life times and decay modes are of fundamental interest in the
physics of the strong interaction. The masses of the pseudo-scalar
mesons have been found to be essentially due to a spontaneous breaking
of the chiral $SU(3)_R\times SU(3)_L$ symmetry or the $U(1)_A$ anomaly
(in case of the $\eta^\prime$). The vector mesons $\rho$, $\omega$, $\phi$,
$K^\star$, $J/\Psi$ etc., which are the dipole modes of the vacuum, have
found increasing attention during the last two decades. Especially
their decay to dileptons is presently investigated in elementary and
complex (nucleus-nucleus) collisions in different laboratories all over
the world (cf. the reviews \cite{Gerry,Cass99,Rapp00} and Refs.
therein). On the other hand, the scalar sector of vacuum excitations is
not well known experimentally and theoretically, so far.

The structure of the lightest scalar mesons $a_0(980)$ and $f_0(980)$
is still under discussion (see e.g.
\cite{Clo,Gen,Jan,Ani,Tornqvist,Hadron99a,Hadron99b} and references
therein). Different authors interpreted them as unitarized $q\bar{q}$
states or as four-quark cryptoexotic states or as $K\bar{K}$ molecules
or even as vacuum scalars (Gribov's minions).  Although it has been
possible to describe them as ordinary $q\bar{q}$-states (see
Refs. \cite{Montanet,Anisovich,Narison}), other options cannot be ruled
out up to now.  Another problem is the possible strong mixing between
the uncharged $a_0(980)$ and the $f_0(980)$ due to a common coupling to
$K\bar K$ intermediate states
\cite{Achasov,Achasov2,Barnes,Speth},\cite{Kerbikov,Close2000,Grishina2001}.
This effect can influence the structure of the uncharged component of
the $a_0(980)$ and implies that it is important to perform a
comparative study of $a_0^0$ and $a_0^+$ (or $a_0^-$).  There is no
doubt that new data on $a_0^0$ and $a_0^+/a_0^-$ production in $\pi N$
and $NN$ reactions are quite important to shed new light on the $a_0$
structure and the dynamics of its production.

In our recent paper
\cite{Grishina} we have considered $a_0$ production in the reaction
$\pi N \to a_0N$ near the threshold and at GeV energies. An effective
Lagrangian approach as well as the Regge pole model were applied to
investigate different contributions to the cross-section of the
reaction  $\pi N \to a_0N$.  Here we employ the latter results for an
analysis of $a_0$ production in $NN$ collisions.  Our study is
particularly relevant to the current experimental program at
COSY-J\"ulich \cite{COSY1,COSY2,COSY3}.

Our paper is organized as follows: In Sect. 2 we discuss an effective
Lagrangian approach with one-pion exchange while the Reggeon exchange
model is considered in Sect. 3.  Sect. 4 is devoted to the calculations
of the cross section for the reaction $NN \to NN a_0$.  In Sect. 5 we
analyze the contribution of the $a_0$ resonance to the cross sections
and invariant $K \bar K$ mass distributions for the reactions $pp \to
pp K^+K^-$ and $pp \to pn K^+ \bar K^0$. Our conclusions are presented
in Sect. 6.

\section{An effective Lagrangian approach with one-pion exchange}

We consider $a_0^0$, $a_0^+$, $a_0^-$ production in the reactions
$j=pp\to pp a_0^0$, $pp\to pn a_0^+$, $pn\to pp a_0^-$ and $pn\to pn
a_0^0$ using the effective Lagrangian approach with one-pion exchange
(OPE). For the elementary $\pi N\to N a_0$ transition amplitude we take
into account different mechanisms $\alpha$ corresponding to $t$-channel
diagrams with $\eta(550)$- and $f_1(1285)$-meson exchanges
($\alpha=t(\eta)$, $t(f_1)$) as well as $s$- and $u$-channel graphs with
an intermediate nucleon ($\alpha=s(N)$, $u(N)$) (cf. Ref. \cite{Grishina}).
The corresponding diagrams are shown in Fig.~\ref{diagr_a0}.  The
invariant amplitude of the $NN\to NN a_0$ reaction then is the sum of
the four basic terms (diagrams in Fig. \ref{diagr_a0}) with
permutations of nucleons in the initial and final states
\begin{eqnarray}
\hspace*{-10mm}{\mathcal{M}}_{j(\alpha)}^{\pi}[ab;cd]
&&=\xi^{\pi}_{j(\alpha)}[ab;cd] \ {\mathcal{M}}_{\alpha}^{\pi}[ab;cd] +
\xi^{\pi}_{j(\alpha)}[ab;dc] \ {\mathcal{M}}_{\alpha}^{\pi}[ab;dc]
\label{NNa0sum} \\
&&+\xi^{\pi}_{j(\alpha)}[ba;dc] \ {\mathcal{M}}_{\alpha}^{\pi}[ba;dc] +
\xi^{\pi}_{j(\alpha)}[ba;cd] \ {\mathcal{M}}_{\alpha}^{\pi}[ba;cd],
\nonumber
\end{eqnarray}
where the coefficients $\xi^{\pi}_{j(\alpha)}$ are given in
Table~\ref{Tab1}.  The amplitude for the $t$-channel exchange with
$\eta(550)$- and  $f_1(1285)$-mesons are given by
\begin{eqnarray}
\hspace*{-10mm}{\mathcal{M}}_{t(\eta)}^{\pi}[ab;cd] &=&
g_{a_0\eta\pi} F_{a_0\eta\pi}\left((p_a-p_c)^2,(p_d-p_b)^2\right)
\  g_{\eta NN}
F_{\eta }\left((p_a-p_c)^2\right)
\nonumber \\
&\times&  {1\over (p_a-p_c)^2-m_\eta ^2} \
\bar u(p_c) \gamma_5 u(p_a)\times {\mathrm{\Pi}}(p_b;p_d),
\label{NN-eta}
\end{eqnarray}
\begin{eqnarray}
\phantom{a}\hspace*{-27mm}  {\mathcal{M}}_{t(f_1)}^{\pi}[ab;cd]&=&
-g_{a_0 f_1\pi} F_{a_0 f_1\pi} \left((p_a-p_c)^2,(p_d-p_b)^2\right)
g_{f_1 NN} F_{f_1}\left((p_a-p_c)^2\right)\nonumber\\
&\times&{1\over (p_a-p_c)^2-m_{f_1}^2}
 \ (p_a-p_c+2 \ (p_b-p_d))_\mu  \left(g_{\mu\nu}-{(p_a-p_c)_\mu
(p_a-p_c)_\nu\over m_{f_1}^2}\right) \nonumber\\
&\times&\bar u(p_c) \gamma_5\gamma_{\nu} u(p_a)\times
 {\mathrm{\Pi}}(p_b;p_d),  \label{NN-f1}
\end{eqnarray}
with
\begin{eqnarray}
\hspace*{-10mm}{\mathrm{\Pi}}(p_b;p_d) =
\frac{f_{\pi NN}}{m_{\pi}}\ F_{\pi }\left((p_b-p_d)^2\right)
(p_b-p_d)_{\beta}
\bar u(p_d) \gamma_5 \gamma_{\beta} u(p_b)\
 \frac{1}{(p_b-p_d)^2-m_{\pi}^2}.
\label{piNN}
\end{eqnarray}
The amplitudes for the $s$- and $u$-channels (lower part of
Fig.~\ref{diagr_a0}) are given as
\begin{eqnarray}
\hspace*{-10mm}{\mathcal{M}}_{s(N)}^{\pi}[ab;cd] &=&
{\mathrm{\Pi}}(p_b;p_d)\
\frac{f_{\pi NN}}{m_{\pi}} F_{\pi }\left((p_d-p_b)^2\right)
\ g_{a_0 NN}\
{ F_{N} \left((p_a+p_b-p_d)^2\right)\over (p_a+p_b-p_d)^2-m_N^2}\nonumber\\
&\times& (p_d-p_b)_{\mu}\
 \bar u(p_c)[(p_a+p_b-p_d)_{\delta}\gamma_{\delta}+m_N]
\gamma_5 \gamma_{\mu} u(p_a)\ ,
\label{NN-s}
\end{eqnarray}
\begin{eqnarray}
\hspace*{-10mm}{\mathcal{M}}_{u(N)}^{\pi}[ab;cd] &=&
{\mathrm{\Pi}}(p_b;p_d)\
\frac{f_{\pi NN}}{m_{\pi}} F_{\pi }\left((p_d-p_b)^2\right)
\ g_{a_0 NN}\ {F_{N}\left((p_c+p_d-p_b)^2\right)
\over (p_c+p_d-p_b)^2-m_N^2} \nonumber\\
&\times& (p_d-p_b)_{\mu}
\ \bar u(p_c)\gamma_5 \gamma_{\mu} [(p_c+p_d-p_b)_{\delta}
\gamma_{\delta}+m_N] u(p_a).
\label{NN-u}
\end{eqnarray}
Here $p_a, p_b$ and $p_c, p_d$ are the four momenta of the initial
and final nucleons, respectively.

The effective Lagrangians involving $a_0$ and $f_1$ mesons were taken
in the following forms:
\begin{eqnarray}
&& {\cal L}_{a_0\eta\pi} = g_{a_0\eta\pi} \ \eta(x) \ \pi(x) \ a_0(x),
    \nonumber\\
&& {\cal L}_{a_0f_1\pi} = g_{a_0f_1\pi} \ \epsilon^{f_1}_{\lambda}(x)
   \ \partial^{\lambda}\pi(x) \ a_0(x),     \label{Lagrangians} \\
&& {\cal L}_{a_0NN}=g_{a_0NN} \ \bar {\Psi}_N(x) \ a_0(x)\ \Psi_N(x),
    \nonumber\\
&& {\cal L}_{f_1NN}=g_{f_1NN} \ \epsilon_{\lambda}^{f_1}(x) \
   \bar {\Psi}_N(x) \ \gamma^{\lambda}\ \Psi_N(x).  \nonumber
\end{eqnarray}
We mostly employ coupling constants and form factors from the
Bonn-J\"ulich potentials (see e.g. Refs. \cite{Holinde,Haidenbauer,Bonnf1}).

The functions $F_i$ in Eqs. (\ref{NN-eta})-(\ref{NN-u})
represent form factors for virtual mesons at the
different vertices $i$ ($i=\pi,\eta,f_1$) and for each vertex they
are taken in the monopole form
\begin{eqnarray}
F_i(t)=\frac{\Lambda_i^2-m_i^2}{\Lambda _i^2-t},
\label{form}
\end{eqnarray}
where $\Lambda_i$ is a cut-off parameter.
For the 'effective' $\pi $ exchange we use the coupling constant
$f_{\pi NN}^2/4\pi =0.08$ and cut-off parameter $\Lambda_{\pi
NN}=1.05\div 1.3$~GeV.  In the case of $\eta$ exchange we take
$g_{\eta NN}^2/4\pi=3$, $\Lambda_{\eta NN}$=1.5 GeV and
$g_{a_0\eta\pi}$=2.46 GeV, which results from the width $\Gamma(a_0 \to
\eta \pi$) = 80 MeV.

The contribution of the $f_1$ exchange is calculated with $g_{f_1
NN}=11.2$, $\Lambda_{f_1 NN}=1.5$~GeV from  Ref.~\cite{Bonnf1} and
$g_{a_0 f_1\pi}$=2.5. The latter value for $g_{a_0 f_1\pi}$ corresponds
to \\ $\Gamma_{tot}(f_1)=24$~MeV and $Br(f_1\to a_0\pi)=34\%$. The
same parameters have been used in our previous study of $a_0$
production in $\pi N\to a_0 N$ and $pp\to da_0^+$ reactions
\cite{Grishina}.

For the form factors at the $a_0 f_1 \pi$ (as well as $a_0 \eta\pi$) vertex
factorized forms are applied following the assumption from
Refs.~\cite{Chung,Nakayama},
\begin{eqnarray}
F_{a_0 f_1 \pi}(t_1,t_2)=F_{f_1 NN}(t_1) \ F_{\pi NN}(t_2),
\label{ff_f1pia0}\end{eqnarray}
where $F_{f_1 NN}(t), F_{\pi NN}(t)$ are taken as in (\ref{form}).

According to different versions of the Bonn potential the coupling
constant $g_{a_0NN}^2/4\pi$ can vary from 1.1075 to
2.67~\cite{Holinde,Bonnf1}.  On the other hand, the unitary model for
meson-nucleon scattering \cite{Feuster} gives a different range for
this constant from 0.0026 to 0.88. In the latter model the $a_0$ only
gives a contribution to the $\pi \eta$ background because there are no
known resonances which decay to $a_0 N$. Since the model is extended
only up to energies $\sqrt{s}\leq 1.9$ GeV, which is below the $a_0$
threshold, the meson-nucleon dynamics is not very sensitive to the
$a_0NN$ coupling.  We note that a small value of $g_{a_0NN}^2/4\pi$
certainly contradicts the experimental values of $Br(p\bar p \to
a_0\pi) =0.69\pm 0.12$ \cite{Smith} and $Br(p\bar p \to a_0\omega)
=0.354\pm 0.028$ \cite{Amsler}, which are quite large (see e.g. Refs.
\cite{Haidenbauer,Bonnf1}).  Having in mind these considerations we
take (as well as in Ref. \cite{Grishina}) the minimal value suggested
by the Bonn potential  $g_{a_0NN} \simeq 3.7$. This value is not very
different from the upper value of 3.33 given by the model of Ref.
\cite{Feuster}.

Another problem is the treatment of a virtual nucleon.
In this case -- instead of the product of two monopole
form factors (at the $a_0NN$ and $\pi NN$ vertices) -- we use
a dipole-like form factor,
\begin{eqnarray}
F_{N}(s) = \frac{\Lambda_{N}^4}{\Lambda_{N}^4+(s-m_N^2)^2},
\label{FN}\end{eqnarray}
which is normalized at $s =m^2$ and has the same asymptotics at large $s$
(positive or negative) as $F_i(s)F_j(s)$.

There are a couple of arguments in favour of using the form factor
(\ref{FN}) for virtual nucleons instead of those which are applied for
virtual mesons. In the $t$-channel graph in elastic $NN$ scattering the
value of $t$ is negative and the monopole form factor $F_{\pi}$ as
given by Eq. (\ref{form}) does not have a singularity in the physical
region and decreases with $t$.  For the $s$-channel graph with a
nucleon exchange in the $\pi N \to a_0 N$ amplitude the value  of $s$
is positive in the physical region and the conventional form factor
$$ \frac{\Lambda_{N}^2 - m_N^2}{\Lambda_{N}^2 -s}$$
may have even a pole in the physical region (this happens for
$\Lambda_N = 2$ GeV, which is used in the Bonn potential for a
virtual $a_0$). This undesirable property is absent in the form factor
(\ref{FN}), where we consider the
cut-off $\Lambda_N$ as a free parameter.  In our previous work \cite{Grishina} we
fixed $\Lambda_N$ in the interval 1.2-1.3 GeV using experimental data
on the differential cross section of the reaction $pp \to d a_0^+$ at
$p_{lab}=3.8 \div 6.3$ GeV$/c$ \cite{BNL73}; in this study we take
$\Lambda_{N} =1.24$~GeV as an average value
(see also the discussion in Section 4).

We recall that the functional form of the nucleon form factor
given by (\ref{FN}) was used in many papers, where meson production
in $\pi N$, $\gamma N$ and $NN$ collisions has been discussed
(see e.g. \cite{Nakayama,Feuster,Pearce,Feuster2,Nakayama2,Titov}
and references therein).

The total cross section for $a_0$ production in the isospin reaction
$j$ is given as the coherent sum of the amplitudes (\ref{NNa0sum}) over
all channels ($\alpha=s(N), u(N), t(f_1), t(\eta)$)
integrated over phase space
\begin{eqnarray}
\sigma_{a_0}^j(s) &=& \int dE_c \ dq_0 \ d{\cos}\theta_q \ d\varphi_q
  \ {1\over 2^9 \pi^4 p_a \sqrt{s}}
\ \left|\sum_\alpha {\mathcal{M}}_{j(\alpha)}^\pi[ab;cd]\right|^2.
\label{sig-tot}
\end{eqnarray}
Here $s=(p_a+p_b)^2$ is the total energy of the $NN$ system squared,
$E_c$ and $q_0$ are the energy of the outgoing nucleon and $a_0$ meson,
respectively. $\theta_q$ is the polar angle of the 3-momentum of the
$a_0$-meson ${\bf q}$ in the cms of the initial nucleons defined as
$\theta_q=\widehat{{\bf q},{\bf p_a}}$, while  $\varphi_q$ is the
azimuthal angle of ${\bf q}$ in the cms.

As shown in the analysis in Ref.~\cite{Grishina} the contribution of
the $\eta$-exchange to the amplitude $\pi N \to a_0 N$ is small.
Note that in Ref. \cite{Baru2} only this mechanism was taken into
account for the reaction $pn \to pp a_0^-$.  Here we also include the
$\eta$-exchange because it might be noticeable in those isospin
channels where a strong destructive interference of $u$- and
$s$-channel terms can occur (see below).


\section{The Reggeon exchange model}

Here as in Ref. \cite{Grishina} we also use the Regge-pole model
for the amplitude $\pi N \to a_0 N$ as developed by Achasov and Shestakov
\cite{Achasov2}.
The $s$-channel helicity amplitudes for the reaction $\pi^-p
\rightarrow a_0^0n$  in this approach can be written as
\begin{eqnarray}
M_{\lambda_2^\prime\lambda_2}(\pi^-p\rightarrow a_0^0n) =
\bar u_{\lambda_2^\prime}(p_2^\prime)\ \left[-A(s,t)
+(p_1+p_1^\prime)_\alpha \gamma_\alpha {B(s,t)\over 2}\right]
\gamma_5 u_{\lambda_2}(p_2),
\label{Reg1}\end{eqnarray}
where the invariant amplitudes $A(s,t)$ and $B(s,t)$ do not contain
kinematical singularities.
The relations between the invariant and s-wave helicity amplitudes
are given by
\begin{eqnarray}
M_{++}= -M_{--}
 =\cos{\theta\over 2} \ \left[A(s,t)\sqrt{-t_{min}}
- B(s,t)\sqrt{-t_{max}s}\right], \label{Reg3}
\end{eqnarray}
\begin{eqnarray}
M_{+-}= M_{-+}
 =\cos{\theta\over 2}\ \left[A(s,t)\sqrt{-t_{max}}
- B(s,t)\sqrt{-t_{min}s}\right],
\label{Reg4} \end{eqnarray}
where $\theta$ is the c.m. scattering angle, while $t_{min}$ and $t_{max}$
are the values of $t$ at $\theta$=0$^o$ and 180$^o$, respectively.

In the model of Ref. \cite{Achasov2}
the $s$-channel helicity amplitudes are
expressed through the $b_1$ and the conspiring $\rho_2$ Regge
trajectories exchange as follows
\begin{eqnarray}
M_{++} = \gamma_{\rho_2}(t)
\exp \left[-i {\pi\over 2} \alpha_{\rho _2}(t)\right]
\left(\frac s{s_0}\right)^{\alpha_{\rho_2}(t)},
\label{Reg3a}\end{eqnarray}
\begin{eqnarray}
M_{+-}&=& \sqrt{(t_{\min }-t)/s_0}\
\gamma_{b_1}(t)
\ i \ \exp \left[-i {\pi\over 2} \alpha_{b_1}(t)\right]
\left(\frac s{s_0}\right)^{\alpha_{b_1}(t)}.
\label{Reg4a}\end{eqnarray}
As in Ref. \cite{Grishina} we take the meson Regge trajectories in
linear form $\alpha_j(t) = \alpha_j(0)+\alpha_j^{\prime}(0)t$ with
$\alpha_{b_1}(0) \simeq -0.37$, $\alpha_{\rho _2}(0) \simeq -0.6$ and
$\alpha_{b_1}^{\prime}(0)$= $\alpha_{\rho _2}^{\prime}(0)=0.9$
GeV$^{-2}$.  The residues are parametrized in a convential way,
$\gamma_{\rho_2}(t)=\gamma_{\rho _2}(0)\ \exp (b_{\rho_2}t)$,
$\gamma_{b_1}(t)=\gamma_{b_1}(0)\ \exp (b_{b_1}t)$;  all
parameters were taken the same as in Ref. \cite{Grishina}.  They
correspond to two fits of the Brookhaven data on $d\sigma/dt$ at 18
GeV/c \cite{Dzierba} found by Achasov and Shestakov \cite{Achasov2}:  a)
with pure $\rho_2$ contribution and b) with combined $\rho_2 + b_1$
contribution.

The invariant amplitude corresponding to the diagram of Fig.
\ref{diagr_re} can be written as
\begin{eqnarray}
{\mathcal{M}}_{Regge}^{\pi}[ab;cd] &=&
\bar u(p_c)\ \left[-A(s,t)
+(p_{a_0}+p_d-p_b)_\alpha \gamma_\alpha {B(s,t)\over 2}\right]
\gamma_5 u(p_a)
\nonumber \\
&\times& \bar u(p_d) \gamma_5 u(p_b)\times {\mathrm{\Pi}}(p_b;p_d).
\label{NN-Regge}
\end{eqnarray}

\section{The reaction $N N\to N N a_0$}

In order to demonstrate the sensitivity of the effective OPE model to
the cut-off parameter $\Lambda_{\pi NN}$ used in the $\pi NN$ vertices
we show in Fig.~\ref{cutoff} the total cross section for the reaction
$pp\to pn a_0^+$ for $u(N)$ and $t(f_1)$ channels as a function of the
excess energy $Q=\sqrt{s}-\sqrt{s_0}$, where $\sqrt{s_0}=m_{a_0}+2m_N$,
calculated for different cut-off parameters.  The dotted lines
correspond to $\Lambda_{\pi NN}=0.8$~GeV, the solid lines show the
result for $\Lambda_{\pi NN}=1.05$~GeV whereas the dashed lines
indicate $\Lambda_{\pi NN}=1.3$~GeV. The results for $\Lambda_{\pi
NN}=1.3$~GeV and 0.8 GeV differ by a factor of $\sim 5$.  For our
subsequent calculation we choose $\Lambda_{\pi NN}=1.05$~GeV while
keeping the uncertainty on $\Lambda_{\pi NN}$ in our 'effective' approach
in mind.

Since we have two nucleons in the final state it is necessary to take
into account their final-state-interaction (FSI), which has some
influence on meson production near threshold. For this purpose we adopt
the FSI model from Ref. \cite{BaruFSI} based on the (realistic) Paris
potential. We use, however, the enhancement factor $F_{NN}(q_{NN})$ --
as given by this model -- only in the region of small relative momenta
of the final nucleons $q_{NN} \leq q_0$, where it is larger than 1.
Having in mind that this factor is rather uncertain at larger $q_{NN}$,
where for example contributions of nonnucleon intermediate states to
the loop integral might be important, we assume that $F_{NN}(q_{NN})
=1$ for $q_{NN} \geq q_0$.

In Fig.~\ref{fsi} we show the FSI effect on the total cross section for
the reactions $pp\to pp a_0^0$ (upper part) and $pp\to pn a_0^+$ (lower
part) for $u(N)$, $s(N)$ and $t(f_1)$ channels. The solid lines show
the calculation without FSI whereas the dashed lines indicate the
results with FSI. As seen from Fig.~\ref{fsi}, the FSI effect is
stronger for $pn$ than for $pp$ in the final state due to the Coulomb
repulsive interaction in the $pp$ system and the isospin dependence of
the $NN$ interaction at small relative momenta.

The results of our calculations for the total cross sections with FSI
for the different isospin reactions are presented in Figs.~\ref{pp_q},
\ref{pn_q} as a function of $Q=\sqrt{s}-\sqrt{s_0}$.  In
Fig.~\ref{pp_q} we show the total cross section for the $pp$ reactions
-- $pp\to pp a_0^0$ (upper part) and $pp\to pn a_0^+$ (lower part),
whereas in Fig.~\ref{pn_q} we display the results for the $pn$
reactions -- $pn\to pp a_0^-$ (upper part) and $pn\to pn a_0^0$ (lower
part).  The solid lines with full dots and with open squares (r.h.s.)
represent the results within the $\rho_2$ and $(\rho_2,b_1)$ Regge
exchange model.  The short dotted lines (l.h.s.) corresponds to the
$t(f_1)$ channel, the dotted lines to the $t(\eta)$ channel, the dashed
lines to the $u(N)$ channel, the short dashed lines to the $s(N)$
channel.  The dashed line in the right upper part of Fig.~\ref{pp_q} is
the incoherent sum of the contributions from $s(N)$ and $u(N)$ channels
($s+u$).

As seen from Figs. \ref{pp_q} and \ref{pn_q}, the $u$- and $s$-channels
give the dominant contribution; the $t(f_1)$ channel is small for all
isospin reactions.  For the reactions $pp\to pn a_0^+$, $pn\to pp
a_0^-$ and $pn\to pn a_0^0$ the Regge exchange contribution (extended
to low energies) becomes important and for the $pn\to pn a_0^0$
reaction this contribution is even dominant near threshold.  For the
$pp\to pp a_0^0$ channel the Regge model predicts no contribution from
$\rho_2$ and $\rho_2,b_1$ exchanges due to isospin arguments (i.e. the
vertex with a coupling of three neutral components of isovectors
vanishes); thus only $s$-, $u$- and $t(f_1)$- channels are plotted in
the upper part of Fig.~\ref{pp_q}.

Here we have to point out the influence of the interference between the
$s$- and $u$-channels. According to the isospin coefficients from the
OPE model presented in Table \ref{Tab1}, the phase (of interference
$\alpha$) between the $s$- and $u$- channels
${\mathcal{M}}_{s(N)}^{\pi}+\exp(-i\alpha){\mathcal{M}}_{u(N)}^{\pi}$ is
equal to zero, i.e. the sign between ${\mathcal{M}}_{s(N)}^{\pi}$ and
${\mathcal{M}}_{u(N)}^{\pi}$ is 'plus'.  The solid lines in
Figs.~\ref{pp_q}, \ref{pn_q} indicate the coherent sum of $s(N)$ and
$u(N)$ channels including the interference of the amplitudes
($s+u+int.$). One can see that for $pp\to pn a_0^+$, $pn\to pp a_0^-$
and $pn\to pn a_0^0$ reactions the interference is positive and
increases the cross section, whereas for the $pp\to pp a_0^0$ channel
the interference is strongly destructive since we have identical
particles in the initial and final states and the contributions of $s$-
and $u$-channels are very similar.

Here we would like to comment  about an extension of the OPE (one-pion-
exchange) model to an OBE (one-boson-exchange) approximation, i.e.
accounting for the exchange of $\sigma, \rho, \omega, ...$ mesons as
well as for multi-meson exchanges.  Generally speaking, the total cross
section of $a_0$ production should contain the sum of all the
contributions:
$$\sigma(NN\to NNa_0) = \Sigma_j \sigma_j,$$
where $j=\pi,\sigma,\rho,\omega...$.  Depending on their cut-off
parameters the heavier meson exchanges might give a comparable
contribution to the total cross section for $a_0$ production. An
important point, however, is that near threshold (e.g. $Q \leq 0.3$ GeV
) the energy behaviour of all those contributions is the same, i.e. it
is proportional to the three-body phase space $ \sigma_j \sim Q^2$
(when the FSI is switched off and the narrow resonance width limit is
taken). In this respect we can consider the one-pion exchange as an
effective one and normalize it to the experimental cross section by
choosing an appropriate value of $\Lambda_{\pi}$.  The most appropriate
choice for $\Lambda_{\pi}$ is about 1$\div$1.3 GeV.  Another question
is related to the isospin of the effective exchange.  As it is known from 
a serious of papers on the reactions $NN\to NNX,
X=\eta,\eta^{\prime},\omega,\phi$ near threshold the  most important
contributions to the corresponding cross sections comes from $\pi$ and
$\rho$ exchanges (see e.g. the review \cite{NakayamaReview} and
references therein). In line with those results we assume here that the
dominant contribution to the cross section of the reaction $NN \to
NNa_0$ comes also from  the isovector exchanges (like $\pi$ and
$\rho$).  In principle, it is also possible that some baryon resonances
may contribute. However, as mentioned above, there is no information
about resonances which couple to the $a_0N$ system.  Our assumptions
thus enable us to make exploratory estimates of the $a_0$ production
cross section without introducing free parameters that would be out of
control by existing data. The model can be extended accordingly when
new data on the $a_0$ production will be available.

Another important question is related to the choice of the form factor
for a virtual nucleon, that -- in line with the Bonn-J\"ulich
potentials -- we choose as given by (\ref{FN}), which corresponds to
monopole form factors at the vertices.  In the literature, furthermore,
dipole-like form factors (at the vertices) are also often used (cf.
Refs. \cite{Nakayama,Feuster,Pearce,Feuster2,Nakayama2}).  However,
there are no strict rules for the 'correct' power of the nucleon form
factor. In physics terms, the actual choice of the power should  not be
relevant; we may have the same predictions for any reasonable choice of
the power if the cut-off parameter $\Lambda_N$ is fixed accordingly.
Note, that $\Lambda_N$ may also depend on the type of mesons involved
at the vertices. Therefore, we can not simply employ the parameters from  
Refs. \cite{Nakayama}, \cite{Pearce} or others in case of the $a_0$ problem.

In our previous work \cite{Grishina} we have fixed $\Lambda_N$ for the
monopole related form factor (\ref{FN})  in the interval 1.2-1.3 GeV
fitting the forward differential cross section of the reaction $pp \to
da_0^+$ from \cite{BNL73}. On the other hand, the same data can be
described rather well using a dipole form factor (at the vertices) with
$\Lambda_N=$1.55-1.6 GeV (cf. Fig. \ref{deutrLam}). If we employ this
dipole form factor with $\Lambda_N=$1.55-1.6 GeV in the present case we
obtain practically identical predictions for the cross sections of the
channels $pp \to pn a_0^+$,  $pn \to pn a_0^0$, $pn \to pp a_0^-$,
where the $u$-channel mechanism is dominant and $u-s$ interference is
not too important. In the case of the channel $pp \to pp a_0^0$ we
obtain cross sections by up to a factor of 2 larger for the dipole-like
form factor in comparison to the monopole one.  This is related to the
strong destructive interference of the $s$ and $u$ exchange mechanisms,
which slightly depends on the type of form factor used.  However, our
central result, that the cross section for the $pn a_0^+$ final channel
is about an order of magnitude higher than the $ppa_0^0$ channel in
$pp$ collisions, is robust (within less than a factor of 2) with
respect to different choices of the form factor.

As seen from Figs.~\ref{pp_q}, \ref{pn_q}, we get the largest cross
section for the $pp\to pn a_0^+$ isospin channel. For this reaction the
$u$-channel gives the dominant contribution, the $s$-channel cross
section is small such that the interference is not so essential as for
the $pp\to pp a_0^0$ reaction.

The result within the Regge model is shown in Fig.~\ref{regge} for the
reactions $pp\to pn a_0^+$, $pn\to pn a_0^0$ (upper part) and $pn\to pp
a_0^-$ (lower part) in a wide energy regime for $Q=$1~MeV$\div$10~GeV.
The total cross section is calculated with the $\rho_2$ (dashed lines)
and $(\rho_2,b_1)$ (solid lines)  Regge trajectories (with FSI) for a
cut-off parameter $\Lambda_{\pi NN}=1.05$~GeV. In order to show the
influence of the cut-off parameter $\Lambda_{\pi NN}$ in the Regge
model we present in the lower part of Fig.~\ref{regge} also the results
for $\Lambda_{\pi NN}=1.3$~GeV (dotted line for $\rho_2$ exchange and
the dot-dashed line for the $(\rho_2,b_1)$ trajectory). Changing the
cut-off $\Lambda_{\pi NN}$ from 1.05 to 1.3 GeV gives a factor $\sim 2$
in the total cross section similar to the results within the effective
Lagrangian model (cf. Fig.~\ref{cutoff}).

As it was already discussed in our previous study \cite{Grishina} an
effective Lagrangian model cannot be extrapolated to high energies
because it predicts the elementary amplitude $\pi N \to a_0N$ to rise
fast. Therefore, such model can only be employed not far from the
threshold; at larger energies it has to be unitarized.  On the other
hand, the Regge model is valid at large energies and we have to worry,
how close to the threshold we can extrapolate corresponding amplitudes.
According to duality arguments one can expect that the Regge amplitude
can be applied at low energy, too, if the reaction $\pi N \to a_0N$
does not contain essential $s$-channel resonance contributions. In this
case the Regge model might give a realistic estimate of the $\pi N \to
a_0N$ amplitude even near threshold.

Anyway, as we have shown in our previous paper \cite{Grishina} the
Regge and $u$-channel model give quite similar results for the $\pi^- p
\to a_0^0 n$ cross-section in the near threshold region; some
differences in the cross sections of the reactions $NN \to NNa_0$ -- as
predicted by those two models -- can be attributed to differences in
the isospin factors and effects of $NN$ antisymmetrization which is
important near threshold (the latter was ignored in the Regge model
formulated for larger energies).

\section{The reaction $N N\to NN a_0 \to N N K \bar K$}

\subsection{The $K\bar K$ and $\pi\eta$ decay channels
of the $a_0(980)$}

The $a_0(980)$ meson production has not yet been measured
in $NN\rightarrow NN a_0$ reactions.  There are only a few
$pp\rightarrow pp K \bar K$ and $pp\rightarrow pn K \bar K$
experimental data points.  Therefore, it is important to analyse a
possible resonance contribution to $K\bar K$ production in the
reactions $NN\rightarrow NN X$, using the calculated $NN\rightarrow NN
a_0$ amplitudes and the experimental fits obtained for the $a_0$
resonance mass distribution in the $K\bar K$ decay channel.

The amplitude for the $a_0(980)$ decays into $K\bar K$ and
$\pi\eta$ modes can be parametrized by the well-known Flatt\'e formula
\cite{Flatte} which satisfies both requirements of analyticity and
unitarity for the two-channels $\pi\eta$ and $K\bar K$.

In the case of the $a_0(980)$ resonance the mass
distribution of the final $K\bar K$ system can be written
as a product of the total cross section for $a_0$ production
(with the 'running' mass $M$) in the $NN\to NN a_0$ reaction
(\ref{sig-tot}) and the Flatt\'e mass distribution
function
\begin{eqnarray}
\frac{d\sigma _{K\bar K}}{d M^2} (s,M) = \sigma_{a_0}(s,M)
\ C_F \frac{M_R \Gamma_{a_0 K\bar K}(M)}
{(M^2-M_R^2)^2
+ M_R^2 \Gamma_{tot}^2(M)} \label{dsdmKK}
\end{eqnarray}
with the total width $\Gamma_{tot}(M)=\Gamma_{a_0 K\bar K}(M)+
\Gamma_{a_0 \pi\eta}(M)$.
The partial widths
\begin{eqnarray}
&&\Gamma_{a_0 K\bar K}(M) = g_{a_0 K\bar K}^2 {q_{K\bar K}\over 8\pi M^2},
\nonumber\\
&&\Gamma_{a_0 \pi\eta}(M) = g_{a_0 \pi\eta}^2 {q_{\pi\eta}\over 8\pi M^2}
\label{width}\end{eqnarray}
are proportional to the decay momenta in the center-of-mass
(in case of scalar mesons),
\begin{eqnarray}
&& q_{K\bar K} = {\left[(M^2-(m_{K}+m_{\bar K})^2)
(M^2-(m_{K}-m_{\bar K})^2)\right]^{1/2} \over 2M} \nonumber\\
&& q_{\pi\eta}={\left[(M^2-(m_{\pi}+m_{\eta})^2)
(M^2-(m_{\pi}-m_{\eta})^2)\right]^{1/2} \over 2M}
\nonumber\end{eqnarray}
for a meson of mass M decaying to  $K\bar K$ and $\pi\eta$, correspondingly.
The branching ratios $Br(a_0\to K\bar K)$ and $Br(a_0\to \pi\eta)$
are given by the integrals of the Flatt\'e distibution
over the invariant mass squared $dM^2 = 2 M dM$:
\begin{eqnarray}
\phantom{a}\hspace*{-3mm}
&&Br(a_0\!\to\! K\bar K)=\!\!\!\!\!\!\!\int\limits_{m_K+m_{\bar K}}^{\infty}
\!\!\!\!\!\!\frac{dM \ 2\ M\ C_F \ M_R\ \Gamma_{a_0 K\bar K}(M)}
{(M^2-M_R^2)^2+M_R^2 \Gamma_{tot}^2(M)},
\label{BrKK}  \\
\phantom{a}\hspace*{-3mm}
&&Br(a_0\!\to\! \pi\eta)=\!\!\!\!\!\!\!\int\limits_{m_K+m_{\bar K}}^{\infty}
\!\!\!\!\!\!\frac{dM \ 2 \ M \ C_F \ M_R \ \Gamma_{a_0 \pi\eta}(M)}
{(M^2-M_R^2)^2+M_R^2 \Gamma_{tot}^2(M)}
\label{Brpieta}\\
\phantom{a}\hspace*{-3mm}
&&\!\!+\int\limits_{m_{\pi}+m_{\eta}}^{m_K+m_{\bar K}}
\!\!\frac{dM \ 2 \ M \ C_F \ M_R \  \Gamma_{a_0 \pi\eta}(M)}
{(M^2-M_R^2-M_R \Gamma_{a_0 K\bar K}(M))^2+M_R^2 \Gamma_{a_0 \pi\eta}^2(M)}.
\nonumber
\end{eqnarray}
The parameters $C_F, g_{K\bar K}, g_{\pi\eta}$ have to be fixed under the
constraint of the unitarity condition
\begin{eqnarray}
Br(a_0 \to K\bar K) + Br(a_0 \to \pi \eta)=1 \ .
\label{unitar}
\end{eqnarray}
Choosing the parameter $\Gamma_0=\Gamma_{a_0 \pi \eta}(M_R)$ in the
interval $50 \div 100$ MeV as given by the PDG \cite{PDG},
one can fix the coupling $g_{\pi\eta}$ according to (\ref{width}).
In Ref. \cite{CrysBar98} a ratio of branching ratios has been reported,
\begin{eqnarray}
r(a_0(980))=\frac{Br(a_0\to K\bar K)}{Br(a_0\to \pi \eta)}=0.23\pm 0.05,
\label{ratBr}
\end{eqnarray}
for $m_{a_0}=0.999$~GeV, which gives $Br(a_0\to K\bar K)=0.187$.
In another recent study \cite{WA102} the WA102 collaboration
reported the branching ratio
\begin{equation}
\Gamma(a_0\to K\bar K) / \Gamma(a_0\to \pi \eta) = 0.166\pm
0.01\pm 0.02 , \label{eq02}
\end{equation}
which was determined from the measured branching ratio for the
$f_1(1285)$-meson.
In our present analysis we use the results from \cite{CrysBar98},
however, keeping in mind that this branching ratio $Br(a_0\to K\bar K)$
more likely gives an 'upper limit' for the $a_0\to K \bar K$ decay.

Thus, the two other parameters in the Flatt\'e distribution $C_F$ and
$g_{a_0 K\bar K}$ can be found by solving the system of integral
equations, for example, Eq. (\ref{BrKK}) for $Br(a_0 \to K \bar K)=0.187$
and the unitarity condition (\ref{unitar}).
For our calculations we choose either $\Gamma_{a_0 \pi \eta}(M_R)=70$~MeV
or 50 MeV, which gives two sets of independent parameters
$C_F, g_{a_0 K \bar K}, g_{a_0 \pi \eta}$ for a fixed branching
ratio $Br(a_0 \to K \bar K)=0.187$:
\begin{eqnarray}
&&\hspace*{-8mm}
set \ 1 \ \ (\Gamma_{a_0\pi\eta}=70~{\rm MeV}): \label{set1}\\
&&\phantom{a}\hfill  g_{a_0 K \bar K}=2.297, \ g_{a_0 \pi \eta}=2.189,
\ C_F=0.365 \nonumber\\
&&\hspace*{-8mm}
set \ 2 \ \ (\Gamma_{a_0\pi\eta}=50~{\rm MeV}): \label{set2}\\
&&\phantom{a}\hfill  g_{a_0 K \bar K}=1.943, \ g_{a_0 \pi \eta}=1.937,
\ C_F=0.354.\nonumber
\end{eqnarray}
Note, that for the $K^+K^-$ or $K^0 \bar K^0$  final state one has
to take into account an isospin factor for the coupling constant, i.e.
$g_{a_0 K^+K^-}=g_{a_0 K^0 \bar K^0} = g_{a_0 K\bar K}/\sqrt{2}$,
whereas $g_{a_0 K^+\bar K^0}=g_{a_0 K^- \bar K^0} = g_{a_0 K\bar K}$.

\subsection{Numerical results for the total cross section}

In the upper part of Fig.~\ref{pp_kk} we display the calculated total
cross section (within parameter $set \ 1$) for the reaction $pp\to pn
a_0^+ \to pn K^+ \bar K^0$ in comparison to the experimental data for
$pp \to pn K^+ \bar K^0$ (solid dots) from Ref.~\cite{LB} as a function
of the excess energy $Q=\sqrt{s}-\sqrt{s_0}$.  The dot-dashed and solid
lines in Fig. \ref{pp_kk} correspond to the coherent sum of $s(N)$ and
$u(N)$ channels with  interference ($s+u+int.$), calculated with a
monopole form of the form factor (\ref{FN}) with $\Lambda_N=1.24$~GeV
and with a dipole form of (\ref{FN}) with $\Lambda_N=1.35$~GeV,
respectively. We mention that the  latter (dipole) result is in better
agreement with the constraints on the near-threshold production of
$a_0$ in the reactions $\pi^-p \to K^-\bar{K^0} p$ and  $\pi^+p \to K^+
\bar {K^0} p $ \cite{Brat_pipKKN}.  In the middle part of
Fig.~\ref{pp_kk} the solid lines with full dots and with open squares
present the results within the $\rho_2$ and $(\rho_2,b_1)$ Regge
exchange model. The short dashed line shows the 4-body phase space
(with constant interaction amplitude), while the dashed line is the
parametrization from Sibirtsev et al. \cite{Sibirtsev1}.  We note, that
the cross sections for parameter $set \ 2$ are similar to $set \ 1$ and
larger by a factor $\sim 1.5$.

In the lower part of Fig.~\ref{pp_kk} we show the calculated total
cross section (within parameter $set \ 1$) for the reaction $pp\to pp
a_0^0 \to pp K^+ K^-$ as a function of $Q=\sqrt{s}-\sqrt{s_0}$ in
comparison to the experimental data. The  solid dots indicate the data
for $pp \to pp K^0 \bar K^0$ from Ref.~\cite{LB}, the open square for
$pp\to pp K^+K^-$ is from the DISTO collaboration~\cite{DISTO} and the
full down triangels show the data from COSY-11 \cite{COSY11}.

For the $pp \to pp a_0^0\to pp K^+K^-$ reaction (as for $pp\to  pp
a_0^0$) there is no contribution from meson Regge trajectories; $s$-
and $u$-channels give similar contributions such that their
interference according to the effective OPE model (line $s+u+int.$) is
strongly destructive (cf.  upper part of Fig.~\ref{pp_q}).  The
$t(f_1)$ contribution (short dotted line) is practically negligible,
while the $t(\eta)$-channel (dotted line) becomes important closer to the
threshold.

Thus our model gives quite small cross sections for $a_0^0$ production
in the $pp\to pp K^+K^-$ reaction which complicates its experimental
observation for this isospin channel.  The situation looks more
promising  for the $pp\to pn a_0^+ \to pn K^+\bar K^0$ reaction since
the $a_0^+$ production cross section is by an order of magnitude larger
than the $a_0^0$ one. Moreover, as has been pointed out with respect to
Fig.~\ref{pp_q}, the influence of the interference is not so strong as
for the $pp\to pp a_0^0 \to pp K^+K^-$ reaction.

Here we stress again the limited applicability of the effective
Lagrangian model (ELM) at high energies. As seen from the upper part of
Fig. \ref{pp_kk}, the ELM calculations at high energies go through the
experimental data, which is not realistic since also other channels
contribute to $K^+\bar K^0$ production in $pp$ reactions (cf. dashed
line from Ref.~\cite{Sibirtsev1}). Moreover, the ELM calculations are
higher than the Regge model predictions which indicates, that the ELM
amplitudes at high energies have to be reggeized or unitarized.

\subsection{Numerical results for the invariant mass distribution}

As follows from the lower part of Fig.~\ref{pp_kk}, the $a_0$
contribution to the $K^+K^-$ production in the $pp\to pp K^+K^-$
reaction near the threshold is hardly seen. With increasing energy the
cross section grows up, however, even at $Q=0.111$~GeV the full cross
section with interference ($s+u+int.$) gives only a few percent
contribution to the $0.11\pm 0.009\pm 0.046\ \mu$b 'nonresonant' cross
section (without $\phi\to K^+K^-$) from the DISTO collaboration
\cite{DISTO}.

To clarify the situation with the relative contribution of $a_0^0$ to
the total $K^+K^-$ production in $pp$ reactions we calculate the
$K^+K^-$ invariant mass distribution for the $pp\to pp K^+K^-$ reaction
at $p_{lab}=3.67$~GeV$/c$, which corresponds to the kinematical
conditions for the DISTO experiment \cite{DISTO}. The differential
results are presented in Fig.~\ref{distf0a0}.  The upper part  shows
the calculation within parameter $set \ 1$, whereas the lower part
corresponds to $set \ 2$.  The dot-dashed lines (lowest curves)
indicate the coherent sum of $s(N)$ and $u(N)$ channels with
interference ($s+u+int.$) for the $a_0$ contribution.  However, one has
to consider also the contribution from the $f_0$ scalar meson, i.e.
the $pp\to pp f_0\to pp K^+K^-$ reaction.  The $f_0$ production in $pp$
reactions has been studied in detail in Ref.~\cite{bratf0}. Here we use
the result from  this previous work \cite{bratf0} and show in
Fig.~\ref{distf0a0} the contribution from the $f_0$ meson calculated
with parameter $set \ A$ from Ref.~\cite{bratf0} as the solid line with
open circles ($f_0$).

We find that when adding the $f_0$ contribution to the phase-space of
nonresonant $K^+K^-$ production (the short dotted lines in
Fig.~\ref{distf0a0}) and the contribution from $\phi$ decays (resonance
peak around 1.02 GeV), the sum (solid) lines almost perfectly describe
the DISTO data.  This means that there is no visible signal for an
$a_0^0$ contribution in the DISTO data according to our calculations
while  the $f_0$ meson gives some contribution to the $K^+K^-$
invariant mass distribution at low invariant masses $M$, that is $\sim
12\%$ of the total 'nonresonant' cross section from the DISTO
collaboration \cite{DISTO}. Thus the reaction $pp \to pn K^+ \bar K^0$
is more promising for $a_0$ measurements as it has been pointed out in
the previous subsection.

For an experimental determination of the $a_0^+$ we present the
invariant mass distribution of $K^+ \bar K^0$ in the reaction $pp \to
pn K^+ \bar K^0$ at different $Q$ (solid lines) in Fig. \ref{dsdM}. The
dashed lines show the invariant mass distributions for 'background'
(i.e. according to phase space with constant interaction amplitude)
under the assumption that the integrals below the solid and dashed
lines are the same for each $Q$.  We see that the shape of the solid
and dashed lines are practically the same for $Q \leq 50 $ MeV.
Noticeable differences between the lines can be found for $Q \geq 100$
MeV.  This means that a separation of the resonance contribution from
the background very close to threshold can be done only in the case
when the background is small or very well known.

\section{Conclusions}

In this work we have estimated the cross sections of  $a_0$ production
in the reactions $pp\to pp a_0^0$, $pp\to pn a_0^+$, $pn\to pp a_0^-$
and $pn\to pn a_0^0$ near threshold and at medium energies.  Using an
effective Lagrangian approach with one-pion exchange we have analyzed
different contributions to the cross section corresponding to
$t$-channel diagrams with $\eta(550)$- and $f_1(1285)$-meson exchanges
as well as $s$ and $u$-channel graphs with an intermediate nucleon.  We
use the same parameters as in our previous paper where we describe
rather well the Berkeley data \cite{BNL73} on the reaction $pp \to d
a_0^+$.

We additionally have considered  the $t$-channel Reggeon exchange
mechanism with parameters normalized to the Brookhaven data for
$\pi^-p\to a_0^-p$ at 18 GeV/c \cite{Dzierba}.  These results have been
used to calculate the contribution of $a_0$ mesons to the cross
sections of the reactions  $pp\to pn K^+\bar K^0$ and $pp\to pp
K^+K^-$.  Due to unfavourable isospin Clebsh-Gordan coefficients as
well as rather strong destructive interference of the $s$- and
$u$-channel contributions  our model gives quite small cross sections
for $a_0^0$ production in the $pp\to pp K^+K^-$ reaction.  However, the
$a_0^+$ production cross section in  the $pp\to pn a_0^+ \to pn K^+\bar
K^0$ reaction should be larger by about an order of magnitude.
Therefore the experimental observation of $a_0^+$ in the reaction
$pp\to pn K^+\bar K^0$ is much more promising than the observation of
$a_0^0$ in the reaction $pp\to pp K^+K^-$.  We note in passing that the
$\pi\eta$ decay channel is experimentally more challenging since, due
to the larger nonresonant background \cite{Mueler01}, the
identification of the $\eta$-meson (via its decay into photons) in a
neutral-particle detector is required.

We have also analyzed invariant mass distributions of the $K \bar K$
system  in the reaction $pp\to pN a_0 \to pN K \bar K$ at different
excess energies $Q$ not far from threshold.  Our analysis of the DISTO
data on the reaction $pp \to pp K^+K^-$ at 3.67 GeV/c has shown that
the $a_0^0$-meson is practically not seen in $d\sigma /dM$ at low
invariant masses, however, the $f_0$-meson gives some visible
contribution.  In this respect the possibility to measure the $a_0^+$
meson in $d\sigma /dM$ for the reaction $pp\to pn K^+\bar K^0$ (or $\to
d K^+\bar K^0$) looks much more promising not only due to a much larger
contribution for the  $a_0^+$, but also due to the absence of the $f_0$
meson in this channel.

Experimental data on $a_0$ production in $NN$ collisions are
practically absent (except of the $a_0$ observation in the reaction
$pp\to dX$ \cite{BNL73}).  Such measurements might give new information
on the $a_0$ structure. According to Atkinson et al. \cite{Atkinson} a
relatively strong production of the $a_0$ (the same as for the
$b_1(1235)$) in non-diffractive reactions can be considered as evidence
for a $q \bar q$ state rather than a $qq \bar q \bar q$ state. For
example the cross section of $a_0$ production in $\gamma p$ reactions
at 25--50 GeV is about 1/6 of the cross sections for $\rho$ and $
\omega$ production.  Similar ratios are found  in the two-body reaction
$pp \to d X$ at 3.8--6.3 GeV/c where $\sigma (pp \to d a_0^+) =(1/4
\div 1/6)\sigma (pp \to d \rho^+)$.

In our case we can compare $a_0$ and $\omega$ production.  Our model
predicts  $\sigma (pp \to pn a_0^+) = 30 \div 70 \mu$b at $Q \simeq 1$
GeV (see Fig. \ref{regge}) which can be compared with $\sigma (pp \to
pp\omega) \simeq  100 - 200 \mu$b at the same $Q$.  If such a large
cross section could be detected this would be a serious argument in
favour of the $q \bar q$ model for the  $a_0$.

To distinguish between the threshold cusp scenario and a resonance
model one can exploit different analytical properties of the $a_0$
production amplitudes in those approaches. In case of a genuine
resonance the amplitude of $\eta \pi$ and $K \bar K$ production through
the $a_0$ has a pole and satisfies the factorization property.  This
implies that the shapes of the invariant mass distributions in the
$\eta \pi$ and $K \bar K$ channels should not depend on the specific
reaction in which the $a_0$ resonance is produced (for $Q \geq
\Gamma_{tot}$). On the other hand, for the threshold cusp scenario the
$a_0$ bump is produced through the $\pi \eta$ final state interaction.
The corresponding amplitude has a square root singularity and in
general can not be factorized (see e.g.  Ref. \cite{BaruFSI} were the
factorization property was disproven for pp FSI in the reaction $pp \to
pp M$).  This implies that for a threshold bump the invariant mass
distributions in the $\eta \pi$ and $K \bar K$ channels are expected to
be different for different reactions and will even depend on
kinematical conditions (i.e. initial energy and momentum transfer) at
the same exess energy, e.g. $Q\simeq 1$ GeV.

\subsection{Acknowledgments}
The authors are grateful to J. Ritman for stimulating discussions and
useful suggestions, to V. Baru for providing the parametrization of the
FSI enhancement factor and to V. Kleber for a careful reading of the
manuscript.



\begin{table*}[t]
\begin{center}
\begin{tabular}{l c c c c }
\hline
Reaction $j$ (mechanism $\alpha$)
& $\xi^{\pi}_{j(\alpha)}[ab;cd]$ & $\xi^{\pi}_{j(\alpha)}[ab;dc]$ &
$\xi^{\pi}_{j(\alpha)}[ba;dc]$ &  $\xi^{\pi}_{j(\alpha)}[ba;cd]$
\\ \hline
$pp\to pp a_0^0 \ (t(\eta),t(f_1))$
& $+1/\sqrt{2}$ & $-1/\sqrt{2}$ & $+1/\sqrt{2}$ & $-1/\sqrt{2}$
\\
$\hphantom{pp\to pp a_0^0 }\ (s(N))$
& $+1/\sqrt{2}$ & $-1/\sqrt{2}$ & $+1/\sqrt{2}$ & $-1/\sqrt{2}$
\\
$\hphantom{pp\to pp a_0^0 }\ (u(N))$
& $+1/\sqrt{2}$ & $-1/\sqrt{2}$ & $+1/\sqrt{2}$ & $-1/\sqrt{2}$
\\
$\hphantom{pp\to pp a_0^0 }\ \mathrm{Regge}$
& $0$ & $0$ & $0$ & $0$
\\
\hline
$pp\to pn a_0^+ \ (t(\eta),t(f_1))$
& $-\sqrt{2}$ & $0$ & $0$ & $+\sqrt{2}$
\\
$\hphantom{pp\to pp a_0^0 }\ (s(N))$
& $0$ & $+\sqrt{2}$ & $-\sqrt{2}$ & $0$
\\
$\hphantom{pp\to pp a_0^0 }\ (u(N))$
& $+2\sqrt{2}$ & $-\sqrt{2}$ & $+\sqrt{2}$ & $-2\sqrt{2}$
\\
$\hphantom{pp\to pp a_0^0 }\ \mathrm{Regge}$
& $-1$ & $+1$ & $-1$ & $+1$
\\
\hline
$pn\to pp a_0^- \ (t(\eta),t(f_1))$
& $+1$ & $-1$ & $0$ & $0$
\\
$\hphantom{pp\to pp a_0^0 }\ (s(N))$
& $-2$ & $+2$ & $-1$ & $+1$
\\
$\hphantom{pp\to pp a_0^0 }\ (u(N))$
& $0$ & $0$ & $+1$ & $-1$
\\
$\hphantom{pp\to pp a_0^0 }\ \mathrm{Regge}$
& $+1/\sqrt{2}$ & $-1/\sqrt{2}$ & $-1/\sqrt{2}$ &
$+1/\sqrt{2}$
\\
\hline
$pn\to pn a_0^0 \ (t(\eta),t(f_1))$
& $-1$ & $0$ & $+1$ & $0$
\\
$\hphantom{pp\to pp a_0^0 }\ (s(N))$
& $-1$ & $-2$ & $+1$ & $+2$
\\
$\hphantom{pp\to pp a_0^0 }\ (u(N))$
& $-1$ & $+2$ & $+1$ & $-2$
\\
$\hphantom{pp\to pp a_0^0 }\ \mathrm{Regge}$
& $0$ & $+\sqrt{2}$ & $0$ & $-\sqrt{2}$
\\
\hline
\end{tabular}
\end{center}
\caption{\label{Tab1} Coefficients in Eq. (\ref{NNa0sum})
for different mechanisms of the $pp\to pp a_0^0$,
$pp \to pp a_0^+$, $pn\to pp a_0^-$ and  $pn \to pn a_0^0$ reactions. }
\end{table*}

\clearpage
\begin{figure}[h]
\phantom{a}\vspace*{1cm}
\centerline{\psfig{figure=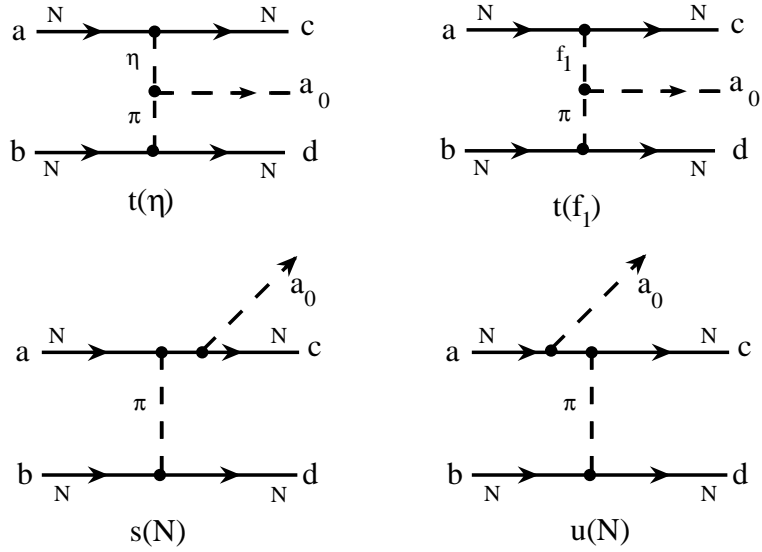,width=10cm}}
\vspace*{5mm}
\caption{Diagrams for $a_0$ production in the reaction
$N N\rightarrow a_0 N N$ near threshold as considered in the present study. }
\label{diagr_a0}
\end{figure}

\begin{figure}[h]
\vspace*{5mm}
\centerline{\psfig{figure=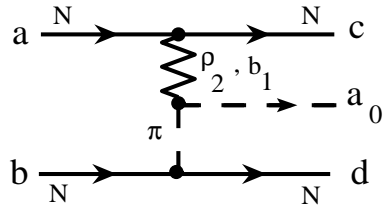,width=5cm}}
\caption{The diagram for $a_0$ production in the reaction
$N N\rightarrow N N a_0$ within the Regge exchange model. }
\label{diagr_re}
\end{figure}

\clearpage
\begin{figure}[h]
\phantom{a}\vspace{3cm}
\centerline{\psfig{figure=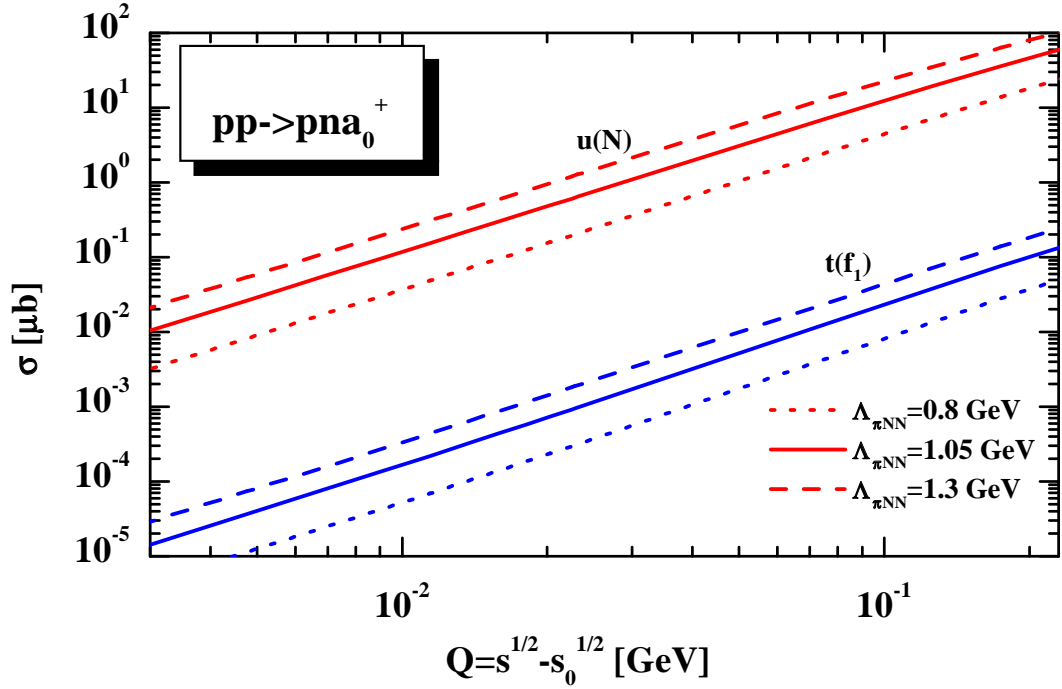,width=14cm}}
\vspace*{5mm}
\caption{The total cross section for the reaction $pp\to pn a_0^+$
for $u(N)$ and $t(f_1)$ channels as a function of the excess energy
$Q=\sqrt{s}-\sqrt{s_0}$ for different cut-off parameters
$\Lambda_{\pi NN}=0.8$~GeV (dotted lines),
$\Lambda_{\pi NN}=1.05$~GeV (solid lines) and
$\Lambda_{\pi NN}=1.3$~GeV (dashed lines).}
\label{cutoff}
\end{figure}

\newpage
\begin{figure}[h]
\phantom{a}\vspace{1cm}
\centerline{\psfig{figure=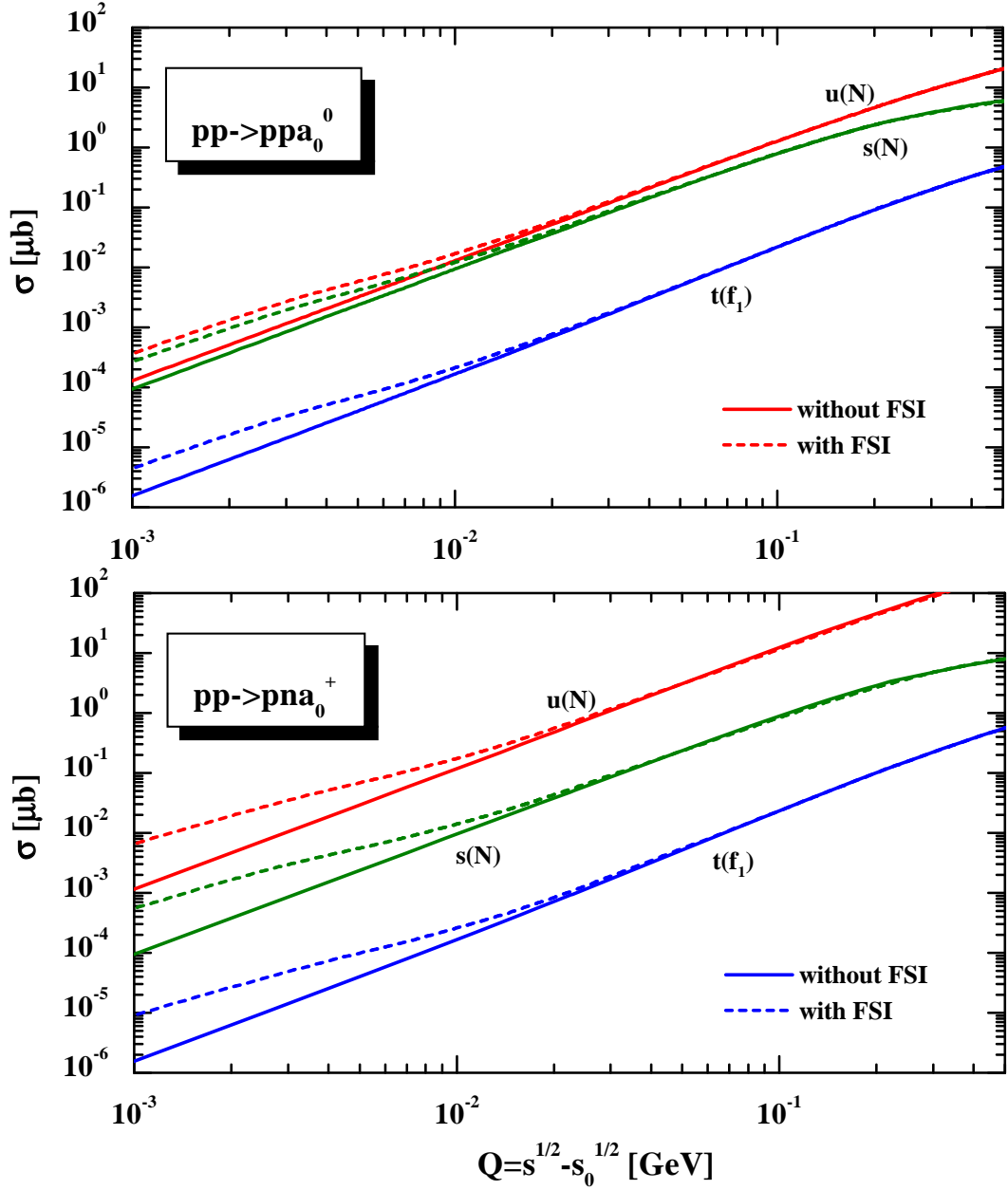,width=14cm}}
\vspace*{5mm}
\caption{The total cross section for the reactions $pp\to pp a_0^0$
(upper part) and $pp\to pn a_0^+$ (lower part) as a function of the
excess energy  $Q=\sqrt{s}-\sqrt{s_0}$ for $u(N)$, $s(N)$
and $t(f_1)$ channels calculated without FSI (solid lines) and with
FSI (dashed lines).}
\label{fsi}
\end{figure}

\newpage
\begin{figure}[h]
\phantom{a}\vspace{1.5cm}
\centerline{\psfig{figure=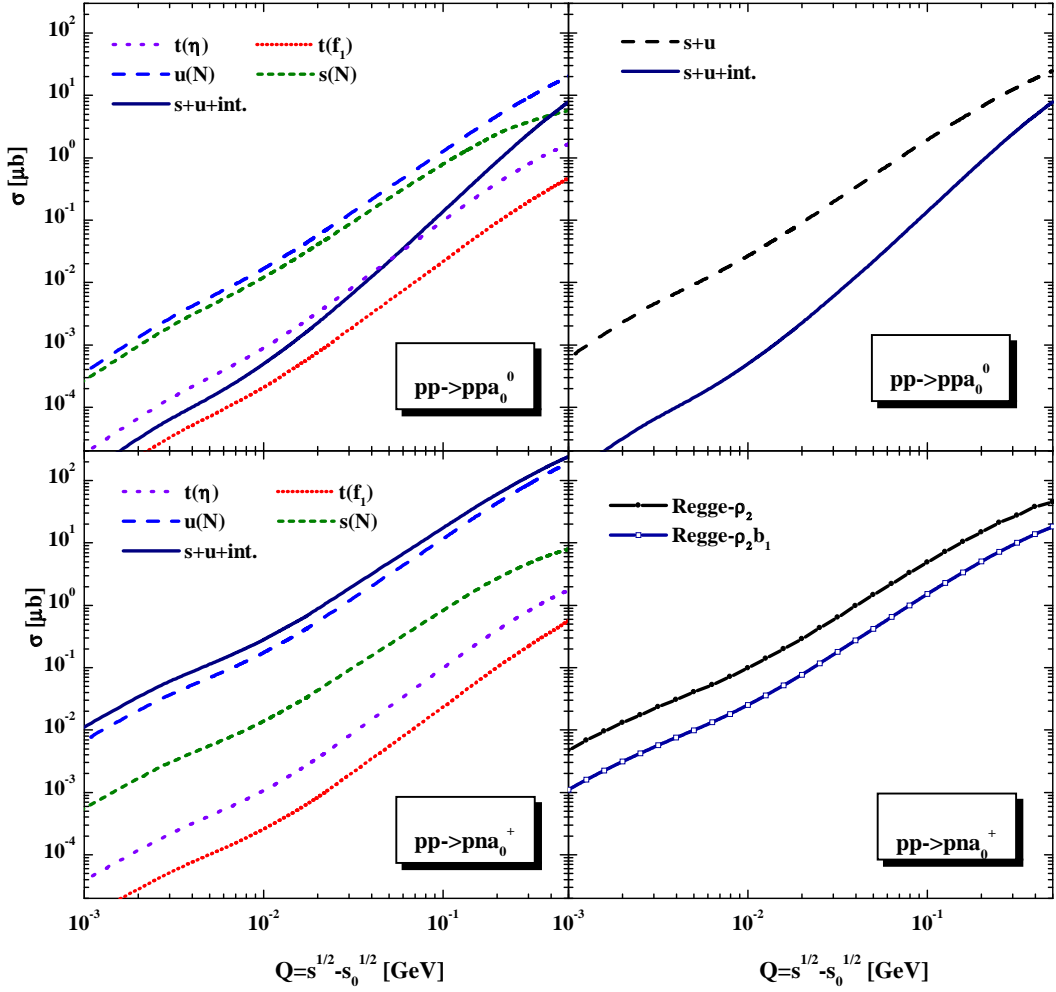,width=14cm}}
\vspace*{5mm}
\caption{The total cross sections for the reactions $pp\to pp a_0^0$
(upper part) and $pp\to pn a_0^+$ (lower part) as a function of the
excess energy $Q=\sqrt{s}-\sqrt{s_0}$ calculated with FSI.
The short dotted lines (l.h.s.) corresponds to the $t(f_1)$ channel,
the dotted lines to the $t(\eta)$ channel,
the dashed lines to the $u(N)$ channel,
the short dashed lines to the $s(N)$ channel.
The dashed line (upper part, r.h.s.) is the incoherent sum of the
contributions from $s(N)$ and $u(N)$ channels ($s+u$).
The solid lines indicate the coherent sum of $s(N)$ and $u(N)$
channels with interference ($s+u+int.$).
The solid lines with full dots and with open squares (lower part,
r.h.s.) present the results within the $\rho_2$ and $(\rho_2,b_1)$
Regge exchange model.}
\label{pp_q}
\end{figure}

\newpage
\begin{figure}[h]
\phantom{a}\vspace{2.5cm}
\centerline{\psfig{figure=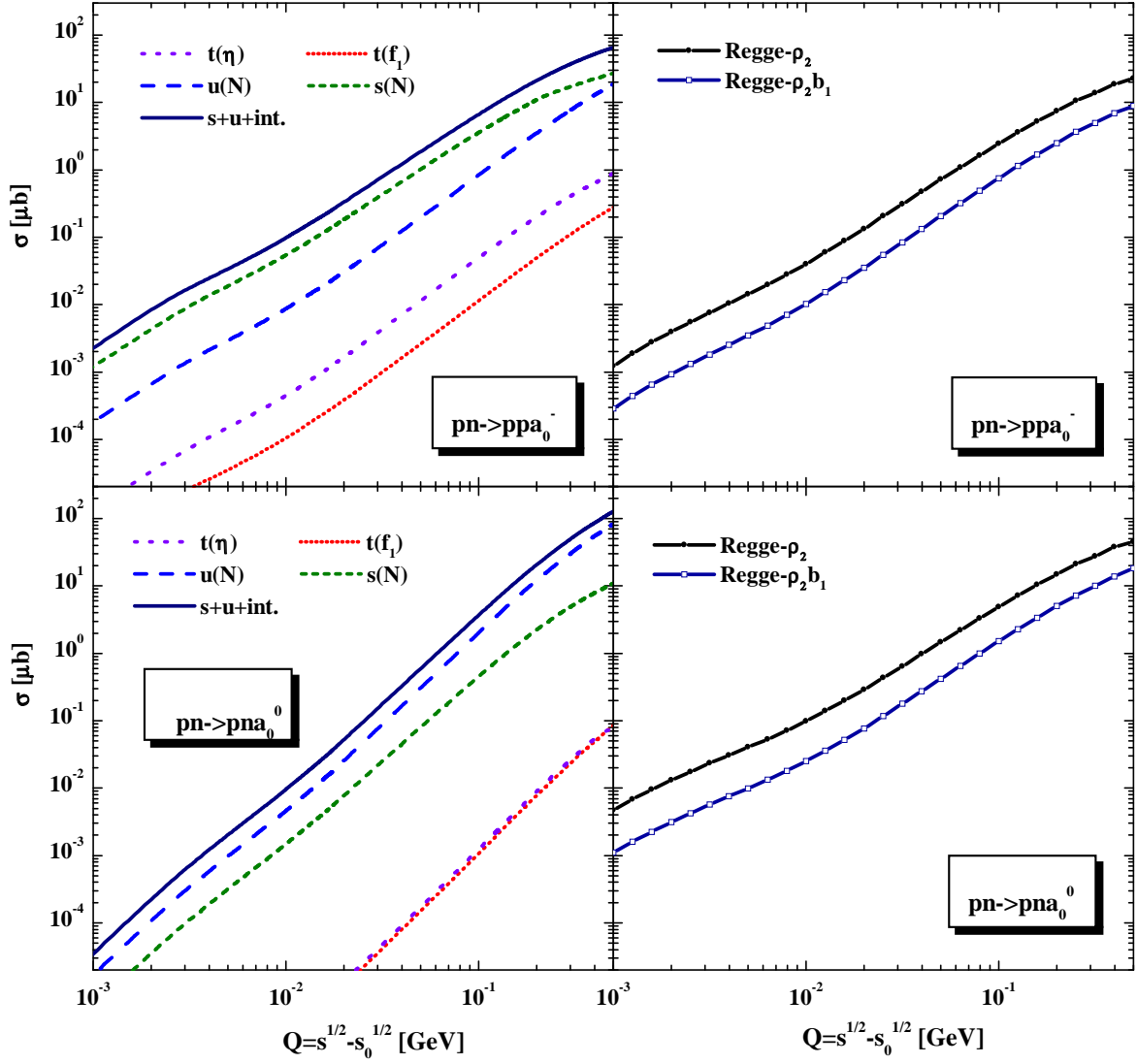,width=15.5cm}}
\vspace*{5mm}
\caption{The total cross sections for the reactions $pn\to pp a_0^-$
(upper part) and $pn\to pn a_0^0$ (lower part) as a function of
$Q=\sqrt{s}-\sqrt{s_0}$ calculated with FSI.  The assignment of the
individual lines is the same as in Fig.  \protect\ref{pp_q}. The
results from the effective OPE model are shown on the l.h.s.  while
those from the $\rho_2$ and $\rho_2 b_1$ Regge approach are displayed on
the r.h.s.}
\label{pn_q}
\end{figure}

\newpage
\begin{figure}[h]
\phantom{a}\vspace{1cm}
\centerline{\psfig{figure=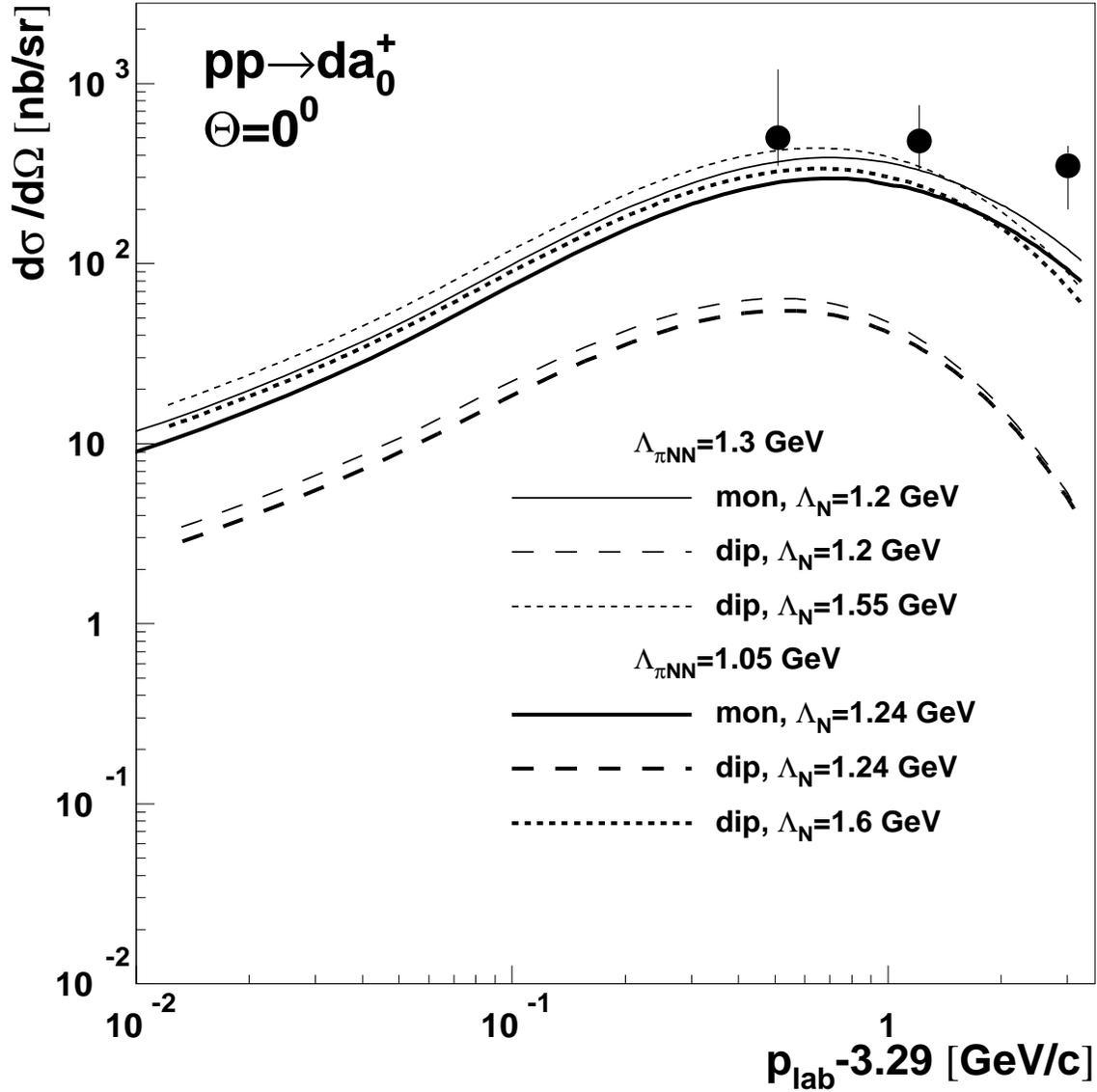,width=15cm}}
\vspace*{5mm}
\caption{Forward differential cross section of the reaction $pp \to d
a_0^+$ as a function of ($p_{lab} -3.29$) GeV/c. The bold and thin
solid curves are calculated at $\Lambda_{\pi NN}$=1.05 and 1.3 GeV,
respectively.  The solid curves correspond to a monopole nucleon form
factor with $\Lambda_N$= 1.2 (thin) and 1.24 GeV (bold). The
long-dashed and short-dashed curves are calculated using the dipole
nucleon form factor for different values of $\Lambda_N$ as shown in the
figure.  The experimental data are taken from
Ref.~\protect\cite{BNL73}. }
\label{deutrLam}
\end{figure}

\newpage
\begin{figure}[h]
\phantom{a}\vspace{1.5cm}
\centerline{\psfig{figure=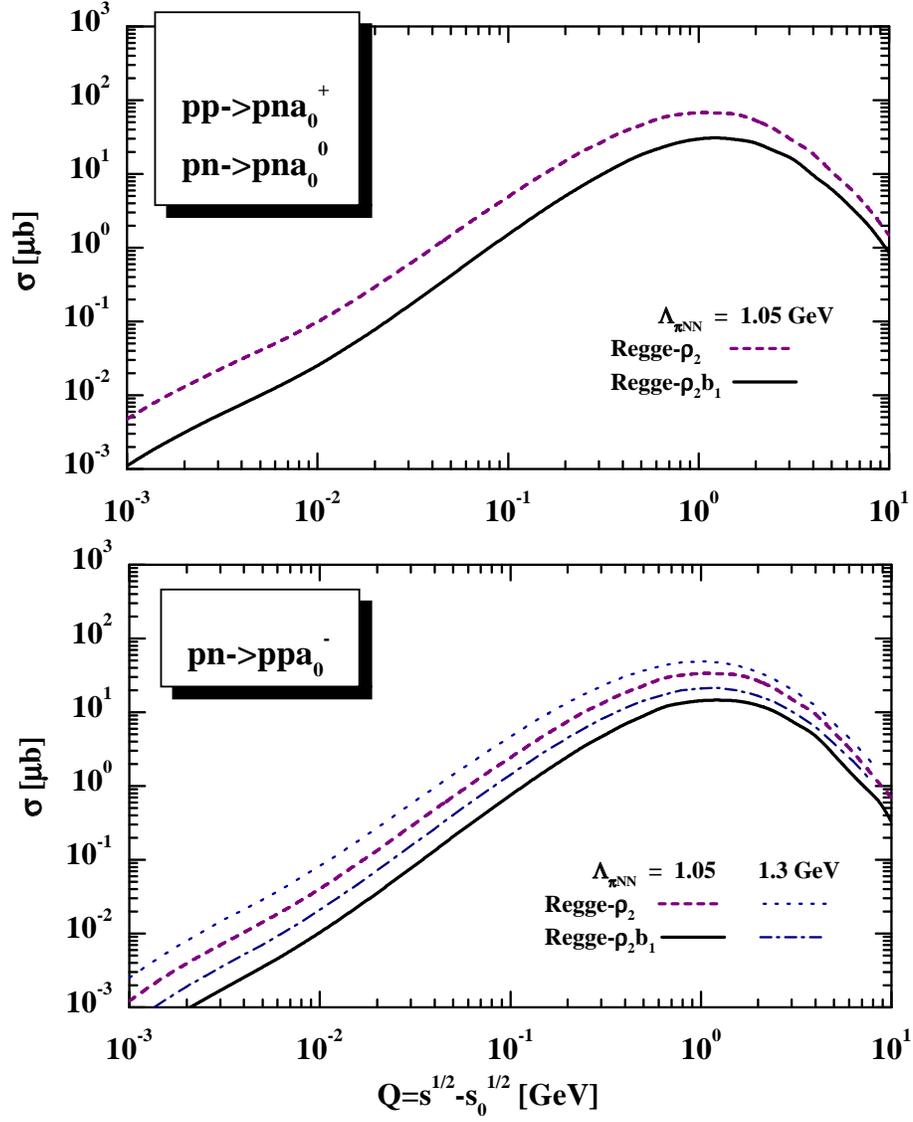,width=12cm}}
\vspace*{5mm}
\caption{The total cross sections for the reactions $pp\to pn a_0^+$,
$pn\to pn a_0^0$ (upper part) and $pn\to pp a_0^-$ (lower part)
as a function of  $Q=\sqrt{s}-\sqrt{s_0}$ calculated
within the $\rho_2$ (dashed lines) and $(\rho_2,b_1)$ (solid lines)
Regge exchange model (with FSI) for cut-off parameters
$\Lambda_{\pi NN}=1.05$~GeV. The dotted and dot-dashed lines
in the lower part show the results for $\Lambda_{\pi NN}=1.3$~GeV
within the $\rho_2$ and $(\rho_2,b_1)$ exchanges, respectively.}
\label{regge}
\end{figure}

\newpage
\begin{figure}[h]
\phantom{a}\vspace{0.5cm}
\centerline{\psfig{figure=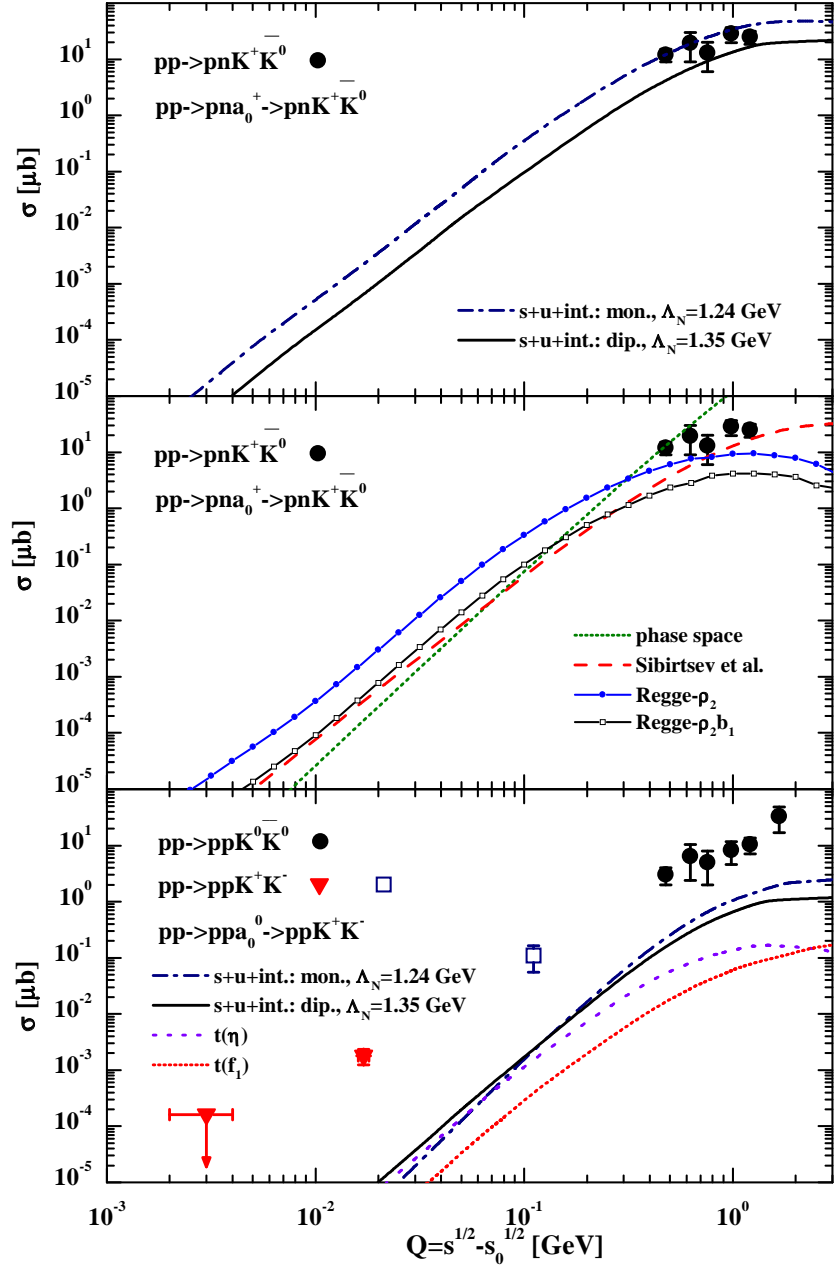,width=11cm}}
\caption{Upper part: the calculated total cross section (within
parameter $set \ 1$) for the reaction $pp\to pn a_0^+ \to pn K^+ \bar
K_0$ in comparison to the experimental data for $pp \to pn K^+ \bar
K_0$ (solid dots) from Ref.~\protect\cite{LB} as a function of
$Q=\sqrt{s}-\sqrt{s_0}$.
The dot-dashed and solid lines  correspond to the coherent sum of
$s(N)$ and $u(N)$ channels with interference ($s+u+int.$)
calculated with a monopole form of the form factor (\protect\ref{FN}) with
$\Lambda_N=1.24$~GeV and with a dipole form of (\protect\ref{FN}) with
$\Lambda_N=1.35$~GeV, respectively.
Middle part: the solid lines with full dots and with open squares
represent the results within the $\rho_2$ and $(\rho_2,b_1)$ Regge
exchange model.  The short dashed line shows the 4-body phase space
(with constant interaction amplitude); the dashed line is the
parametrization from Sibirtsev et al.~\protect\cite{Sibirtsev1}.
Lower part: the calculated total cross section (within parameter $set \
1$) for the reaction $pp\to pp a_0^0 \to pp K^+ K^-$ as a function of
$Q=\sqrt{s}-\sqrt{s_0}$ in comparison to the experimental data.  The
solid dots indicate the data for $pp \to pp K_0 \bar K_0$ from
Ref.~\protect\cite{LB}, the open square for $pp\to pp K^+K^-$ from
Ref.~\protect\cite{DISTO}; the full down triangels show the data from
Ref.~\protect\cite{COSY11}. }
\label{pp_kk}
\end{figure}

\newpage
\begin{figure}[h]
\phantom{a}\vspace{1.5cm}
\centerline{\psfig{figure=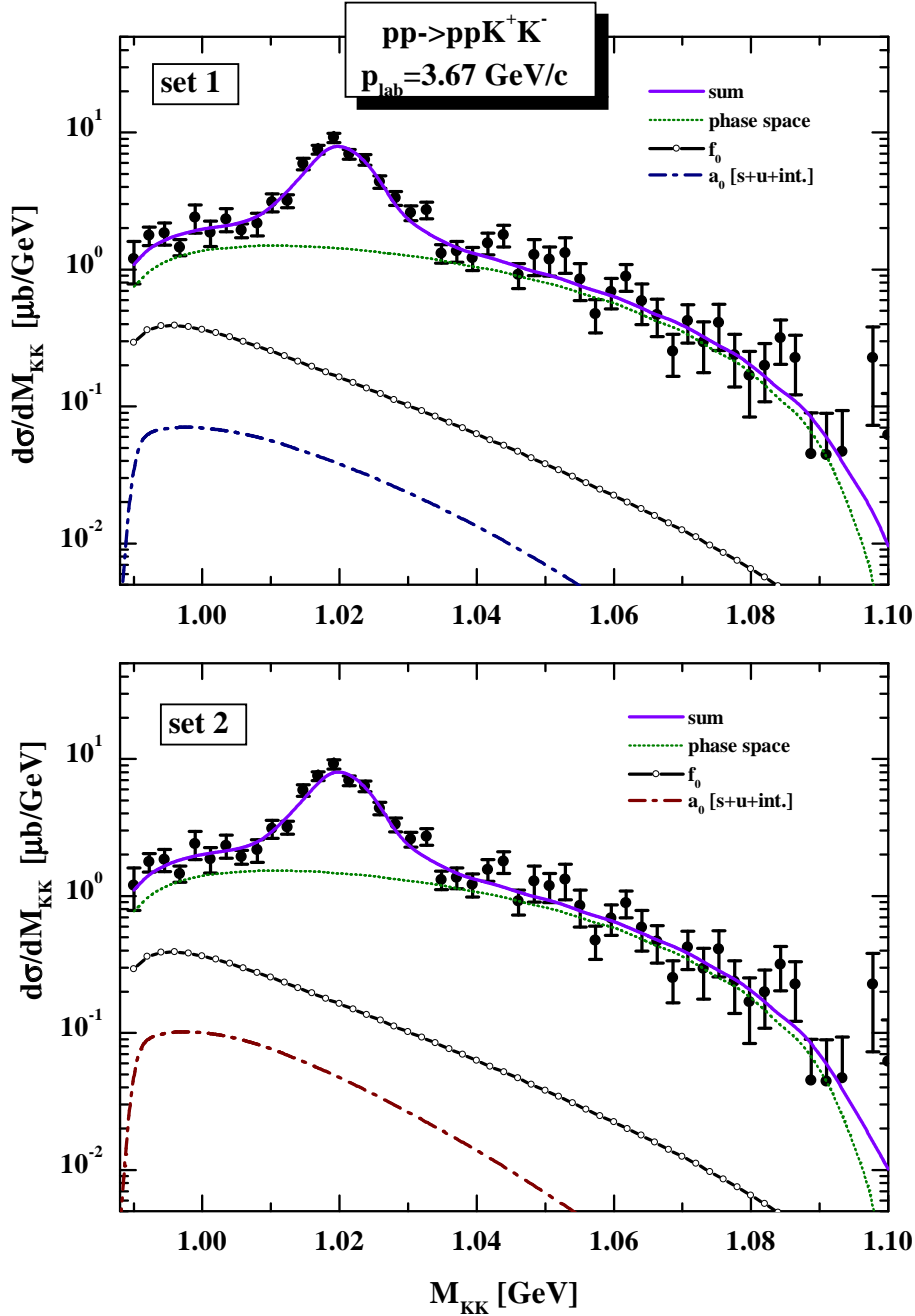,width=12cm}}
\vspace*{5mm}
\caption{The $K^+K^-$ invariant mass distribution for the $pp\to
pp K^+K^-$ reaction at $p_{lab}=3.67$~GeV$/c$.
The short dotted lines indicate the 4-body phase space with constant
interaction amplitude, the dot-dashed lines show the coherent sum
of $s(N)$ and $u(N)$ channels with interference ($s+u+int.$).  The
solid lines with open circles correspond to the $f_0$ contribution from
Ref.~\protect\cite{bratf0}.  The thick solid lines show the sum of all
contributions including the decay $\phi\to K^+K^-$.
The experimental data are taken from Ref.~\protect\cite{DISTO}. }
\label{distf0a0}
\end{figure}

\newpage
\begin{figure}[h]
\phantom{a}\vspace{1.5cm}
\centerline{\psfig{figure=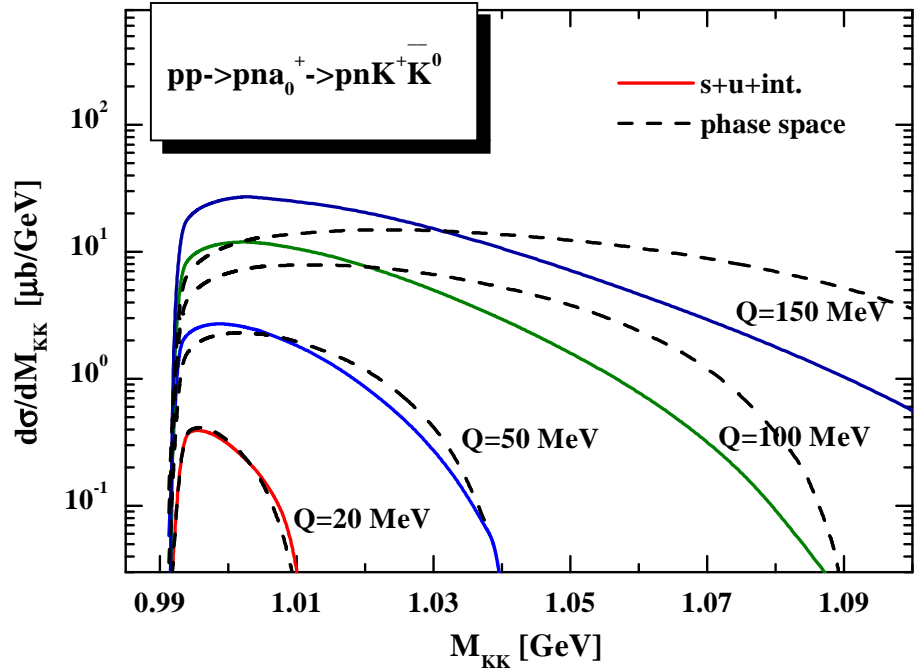,width=12cm}}
\vspace*{5mm}
\caption{The $K^+\bar K^0$ invariant mass distribution for the $pp\to
pn K^+\bar K^0$ reaction at different $Q = \sqrt{s} - \sqrt{s}_0$.  The
solid lines describe the $a_0^+$ resonance contributions.  The dashed
lines show the invariant mass distributions for 'background' under the
assumption that the integrals below the solid and dashed lines are the
same for each $Q$.}
\label{dsdM}
\end{figure}

\end{document}